\documentclass[a4paper, oneside, twocolumn, notitlepage, 10pt]{extarticle_ecoc}
\usepackage{ecoc}

\begin{document}
\selectlanguage{english}    


\title{On the Optimum Energy-per-bit Launch Power in Coherent Hollow-core Fibre Transmission Systems}%


\author{
    Ronit~Sohanpal, Eric~Sillekens, Mindaugas~Jarmolovi\v{c}ius, Robert~I.~Killey and Polina~Bayvel
}

\maketitle                  


\begin{strip}
    \begin{author_descr}

        Optical Networks Group, UCL (University College London), London, UK, 
        \textcolor{blue}{\uline{ronit.sohanpal@ucl.ac.uk}}\\

    \end{author_descr}
\end{strip}

\renewcommand\footnotemark{}
\renewcommand\footnoterule{}


\begin{strip}
    \begin{ecoc_abstract}
        We investigate the optimum energy per bit in hollow-core-fibre transmission systems. We show that a 1000~km C-band link can achieve a 41.5\% reduction in total power consumption when operating at the minimum energy-per-bit launch power with only 2.2\% throughput penalty. ©2026 The Author(s)
    \end{ecoc_abstract}
\end{strip}


\section{Introduction}
Hollow-core fibres (HCFs) have seen a significant research focus in recent years to enhance the performance of optical communication links. HCFs offer considerable advantages over widely-used silica single-mode fibres (SMFs), such as a 30\% reduction in latency, three to four orders of magnitude reduction in nonlinearity and the reduction in loss to below 0.1~dB/km. The reduction in both nonlinearity and attenuation enables a considerable increase in the optimal launch power, which can reach and exceed 40~dBm depending on the HCF attenuation \cite{poggioliniOpportunitiesChallengesLongDistance2022}. This allows for much longer span lengths exceeding 200~km, reducing the total number of link amplifiers to lower deployment complexity and improve energy efficiency \cite{poggioliniPotentialSpanLength2023a,sticcaHighPowerOpticalAmplification2024}. 

Simultaneously, considerable research has focused on exploring ultra-wideband (UWB) technologies to enhance the throughput of single-fibre systems by leveraging the low-loss window of silica, with recent demonstrations making use of over 42~THz of optical bandwidth across the OESCL-band spectrum \cite{yangChallengesBreakthroughsModeling2026,luis450TbGMI2026}. Due to the tunability of HCF designs to achieve low attenuation in various wavelength windows (e.g. O- to U-bands as well as 1~\si{\micro\metre}), UWB HCF transmission is of interest to further scale single-fibre throughputs \cite{petrovichBroadbandOpticalFibre2025,mahdirajiFirstDualBandHybrid2026}. 

However, the power conversion efficiency (PCE) of thulium- and bismuth-doped fibre amplifiers (TDFAs and BDFAs) used for the O-, E- and S-bands is considerably lower than that of the erbium-doped fibre amplifiers (EDFAs) for the C- and L-bands, leading to worse system energy-per-bit as more bands are deployed \cite{sohanpalMeasurementAnalysisPower2025,sohanpalUltraWidebandTransmissionSystems2026}. In addition, in conventional silica fibre links it is generally assumed that for few spans the power consumption of the optical amplifiers is negligible due to the dominant contribution of the numerous coherent transceivers, including in the C-band. As HCFs support considerably higher launch powers than silica fibres (with experimental demonstrations exceeding 34~dBm total launch power \cite{mardoyan222kmlongHybridSpan2026,hongTransmitterPowerOptimization2026}) the validity of this assumption becomes questionable. 

In this work, we investigated the energy efficiency of HCF transmission systems at different launch powers. We show that in HCF, unlike SMF, the amplifier power consumption becomes comparable to the transceiver power consumption at high launch powers, even for few-span systems. This leads to two different launch power operating points, achieving either minimum energy per bit or maximum throughput. Our results showed that for a 1000~km C-band HCF link, operating at the launch power achieving minimum energy per bit can reduce link power consumption by 41.5\% for only a 2.2\% throughput penalty versus maximum-throughput operation, increasing to 71.9\% power reduction for 8.5\% throughput penalty in OESCL-band systems.

\section{Methodology}

\renewcommand{\THz}[1]{\fpeval{299792.458/(#1)}}
    
\begin{figure}[b]

\begin{tikzpicture}

    \begin{axis}
    [
    legend columns=1,
    width=0.4\textwidth,height=5cm,
    grid=both,
    legend style={fill opacity=1, draw opacity=1, text opacity=1, at={(0.5,0.3)}, anchor=south west, draw=black, nodes={scale=0.7, transform shape}},
    xlabel=Frequency (THz),
    ylabel=Attenuation (dB/km),
    ylabel near ticks,
    ylabel shift = -2 pt,
    xlabel shift = -2 pt,
    xticklabel shift = 2 pt,
    ytick = {0.07,0.09,0.11,0.13,0.15},
    yticklabels={$0.07$,$0.09$,$0.11$,$0.13$,$0.15$},
    ymin=0.07,
    ymax=0.15,
    xmin=186,
    xmax=237,
    xtick={\THz{1265},\THz{1355},\THz{1400},\THz{1460},\THz{1530},186},
    xticklabels={$237$,$221$,$214$,$205$,$196$,$186$},
    clip marker paths=true,
    ]

    \newcommand\xaa{0.05}
    \newcommand\xbb{0.2}

    \fill[o-band,opacity=.2] (axis cs:\THz{1265},\xaa) rectangle (axis cs:\THz{1355},\xbb);
    \fill[e-band,opacity=.2] (axis cs:\THz{1400},\xaa) rectangle (axis cs:\THz{1460},\xbb);
    \fill[s-band,opacity=.2] (axis cs:\THz{1460},\xaa) rectangle (axis cs:\THz{1530},\xbb);
    \fill[c-band,opacity=.2] (axis cs:\THz{1530},\xaa) rectangle (axis cs:\THz{1565},\xbb);
    \fill[l-band,opacity=.2] (axis cs:\THz{1570},\xaa) rectangle (axis cs:\THz{1620},\xbb);
    
    \node[anchor=mid] at (axis cs:\THz{1310},0.14){O};
    \node[anchor=mid] at (axis cs:\THz{1430},0.14){E};
    \node[anchor=mid] at (axis cs:\THz{1495},0.14){S};
    \node[anchor=mid] at (axis cs:\THz{1547.5},0.14){C};
    \node[anchor=mid] at (axis cs:\THz{1595},0.14){L};

    \addplot[Set1-B,thick] table[x=freq,y=att] {Data/DispAtt.txt};

    \draw[black,thick] (axis cs: 208.5,0.085) ellipse (0.3cm and 0.15cm);

    \draw [stealth-,line width = 1] (axis cs:200,0.085)--(axis cs:205.5,0.085);

    \end{axis}
   
    \begin{axis}
    [
    legend columns=1,
    width=0.4\textwidth,height=5cm,
    legend style={fill opacity=1, draw opacity=1, text opacity=1, at={(0.5,0.3)}, anchor=south west, draw=black, nodes={scale=0.7, transform shape}},
    xlabel=Frequency (THz),
    ylabel=Dispersion (ps/(nm$\cdot$km),
    ylabel near ticks,
    xlabel near ticks,
    ylabel shift = -2 pt,
    xlabel shift = -2 pt,
    hide x axis,
    axis y line*=right,
    ymin=1.5,
    ymax=3.5,
    xmin=186,
    xmax=237,
    clip marker paths=true,
    ]

    \addplot[Set1-E,thick] table[x=freq,y=D] {Data/DispAtt.txt};

    \draw[black,thick] (axis cs: 214,2.65) ellipse (0.3cm and 0.15cm);

    \draw [-stealth,line width = 1] (axis cs:217.2,2.65)--(axis cs:222.7,2.65);
    
    \end{axis}

\end{tikzpicture}

\caption{Dispersion and attenuation profile based on simulated 1st window DNANF HCF from \cite{mahdirajiFirstDualBandHybrid2026,petrovichBroadbandOpticalFibre2025}.}
\label{fig:1}

\end{figure}
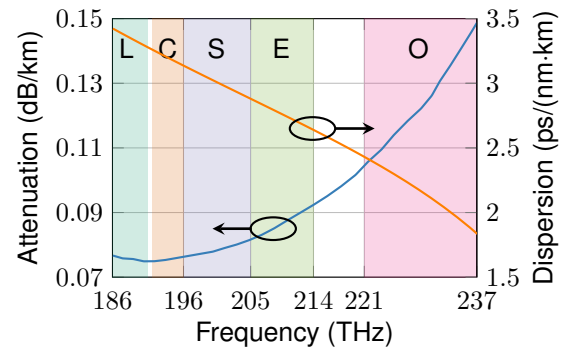

To investigate the energy-per-bit, we used the closed-form GN model to estimate the per-channel SNR. Our approach extends that in \cite{poggioliniOpportunitiesChallengesLongDistance2022}, by including the wavelength-dependence of the HCF attenuation and dispersion to extend the model for UWB systems, shown in Fig.~\ref{fig:1}. The attenuation and dispersion profiles are based on the simulated 1st window DNANF profiles in \cite{petrovichBroadbandOpticalFibre2025,mahdirajiFirstDualBandHybrid2026}, corresponding to an attenuation and dispersion of approximately 0.075~dB/km and 3.2~ps/(nm$\cdot$km) at 1550~nm and 0.12~dB/km and 2.17~ps/(nm$\cdot$km) at 1310~nm. This fibre profile provides low attenuation across the entire OESCL transmission window, reflecting an ideal scenario in terms of minimal absorption impairments (e.g. water vapour, $\mathrm{CO_2}$ and $\mathrm{NO_x}$) which can be mitigated via optimisation of fabrication processes \cite{xiongTextCO_2EliminationHollowCore2025}.

Assuming additive white Gaussian noise (AWGN) and Gaussian modulation, the SNR for a given channel $i$ can be calculated as:
\begin{equation}
    \text{SNR}_i = \frac{P_i}{P_{\text{ASE}} + \eta P_i^3 + P_{\text{IMI}} + P_{\text{TRN}}}
\end{equation}
where $P_i$ is the $i$-th channel launch power, $P_{\text{ASE}}$ is the accumulated ASE noise power, $\eta$ is the nonlinear interference coefficient, $P_{\text{IMI}}$ is the HCF inter-modal interference (IMI) coefficient and $P_{\text{TRN}}$ is the transceiver noise power. We assume that inter-channel stimulated Raman scattering can be neglected. Transceiver noise was included here as a limit to the maximum achievable throughput. 

The single-channel throughput is given by the Shannon capacity, $T_i = 2R\log_2(1+\text{SNR}_i)$ for channel baud rate $R$.

\renewcommand{\THz}[1]{\fpeval{299792.458/(#1)}}
    
\begin{figure}[h]

\begin{tikzpicture}

    \begin{axis}
    [
    legend columns=1,
    width=\linewidth,height=4.3cm,
    grid=major,
    minor tick num=1,
    legend style={fill opacity=1, draw opacity=1, text opacity=1, at={(0.73,0.1)}, anchor=south west, draw=black, nodes={scale=0.7, transform shape}},
    xlabel=Launch power per channel (dBm),
    ylabel= Throughput (Tbps),
    ylabel near ticks,
    ylabel shift = -2 pt,
    xlabel shift = -2 pt,
    xticklabel shift = 2 pt,
    ytick = {1, 10, 60},
    yticklabels={$1$,$10$,$60$},
    ymin=1,
    ymax=80,
    xmin=-40,
    xmax=50,
    ymode=log,
    clip marker paths=true,
    ]

    \addplot[c-band,thick] table[x=lpch,y=capc] {Data/Fig2ThroughputC_HCF.txt};
    \addplot[l-band,thick] table[x=lpch,y=capc] {Data/Fig2ThroughputC_SMF.txt};
    \legend{HCF,SMF};
    
    \addplot[l-band, only marks,mark=square*] coordinates {(10,37.6708621827976)};
    \node[below] at (axis cs:9.5,34) {$T_{\text{tot}}^{\text{max}}$};

    \addplot[c-band, only marks,mark=square*] coordinates {(22,52.6708660350891)};
    \node[below] at (axis cs:22,52.6708660350891) {$T_{\text{tot}}^{\text{max}}$};

    \end{axis}

\end{tikzpicture}

\caption{C-band throughput versus total launch power for 1x200~km HCF and 3x67~km low-loss SMF (200~km total).}
\label{fig:2}

\end{figure}
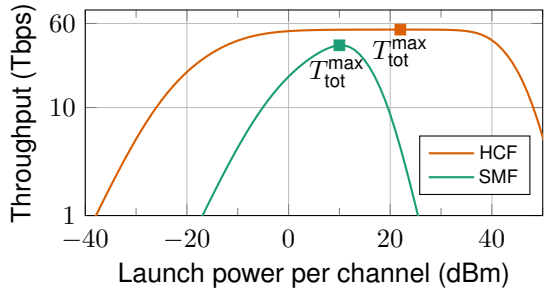

\begin{figure*}[hb!]
\begin{tikzpicture}[font=\footnotesize]
    \begin{groupplot}
    [
    width=0.53\linewidth,
    height=7.5cm,
    grid=major,
    ylabel near ticks,
    xlabel near ticks,
    ylabel shift = -8 pt,
    xlabel shift = -2 pt,
    clip marker paths=true,
    ymode=log,
    xmode=log,
    xlabel=Throughput (Tbps),
    ylabel=Energy per bit (pJ/bit),
    clip mode=individual,
    label style={font=\footnotesize},
    tick label style={font=\footnotesize},
    legend style={fill opacity=1, draw opacity=1, text opacity=1, at={(0.5,0.97)}, anchor=north west, draw=black, nodes={scale=0.75, transform shape}, legend columns=1},
    group style={group size=2 by 1,xlabels at=edge bottom,ylabels at=edge left,horizontal sep=1cm},
    title style={at={(0.5,0.95)},font=\bfseries\footnotesize},
    ]

    \nextgroupplot[
    domain = 5:50,
    xtick={5,10,20,30,40},
    xticklabels={$5$,$10$,$20$,$30$,$40$},
    ytick = {0.1,1,10,100,500},
    yticklabels={$0.1$,$1$,$10$,$100$,$500$},
    ymin=0.1,
    ymax=500,
    xmin=5,
    xmax=40,
    title={\textbf{(a)} SMF, 3x67~km},
    ]

    \node[rotate=-11] at (axis cs: 7,66) {\normalsize $E_b \propto 1/T_{\text{tot}} $};
    
    \addplot[black,
    no marks, 
    dashed,
    line width = 0.8pt,
    ]plot {(29*24)/\x};    

    \addplot[l-band,
    no marks,  
    line width=1.2pt,
    postaction={decorate, decoration={
            markings,
            mark=at position 0.03 with {\arrow{stealth}},
            mark=at position 0.29 with {\arrow{stealth}},
            mark=at position 0.35 with {\arrow{stealth}},
            mark=at position 0.44 with {\arrow{stealth}},
        }}
    ] table[x=capc,y=pjbittrx_c] {Data/Fig3dataSMF_3span.txt};
    
    \addplot[l-band,
    mark=*, 
    only marks,
    mark size = 1.5, 
    line width=0.7pt,
    each nth point=2,
    ] table[x=capc,y=pjbittrx_c] {Data/Fig3dataSMF_3span.txt};

    \addplot[l-band, only marks, 
    mark=square,
    mark size = 1.5, 
    line width=1pt,
    each nth point=2,
    ] table[x=capc,y=pjbitamp_c] {Data/Fig3dataSMF_3span.txt};

    \draw [-stealth,line width = 1] (axis cs: 11,44)--(axis cs:19,27);
    \node[rotate=-10] at (axis cs: 14,24) {Increasing $P_i$};
    
    \draw[black!90,densely dashed] (32,35) rectangle (39,12);

    \node at (axis cs: 23.6, 29.62-8) {\footnotesize $2$};
    \node at (axis cs: 27.96, 25.07-7) {\footnotesize $4$};
    \node at (axis cs: 23.26, 33.44+12) {\footnotesize $16$};
    \node at (axis cs: 15.25, 54.17+20) {\footnotesize $18$};
    \node at (axis cs: 8.64, 104.4+40) {\footnotesize $20$};

    \node at (axis cs: 23.6*1.06, 0.148*0.9) {\footnotesize $2$};
    \node at (axis cs: 27.96*1.06, 0.189*0.9) {\footnotesize $4$};
    \node at (axis cs: 23.26, 3.532*0.7) {\footnotesize $16$};
    \node at (axis cs: 15.25, 8.53*0.7) {\footnotesize $18$};
    \node at (axis cs: 8.64, 23.88*0.7) {\footnotesize $20$};

    \addlegendimage{black,mark=*,
    postaction={decorate, decoration={
            markings,
            mark=at position 0.3 with {\arrow{stealth}},
            mark=at position 0.9 with {\arrow{stealth}}
        }}}
    \addlegendimage{black,mark=square, only marks}
    \legend{,,,,Amplifiers+transceivers,Amplifiers only}



    \nextgroupplot[
     xtick={30,35,40,45,50,55},
    xticklabels={$30$,$35$,$40$,$45$,$50$,$55$},
    ytick = {0.01,0.1,1,10,100,1000,10000},
    yticklabels={$0.01$,$0.1$,$1$,$10$,$100$,$1000$,$10000$},
    ymin=0.01,
    ymax=1000,
    xmin=30,
    xmax=55,
    domain = 30:60,
    title={\textbf{(b)} HCF, 1x200~km},
    ]

    
    \addplot[black,
    no marks, 
    dashed,
    line width = 0.8pt,
    ]plot {(29*24)/\x};    


    \addplot[c-band,
    no marks,  
    line width=1.2pt,
     postaction={decorate, decoration={
            markings,
            mark=at position 0.03 with {\arrow{stealth}},
            mark=at position 0.08 with {\arrow{stealth}},
            mark=at position 0.2 with {\arrow{stealth}},
            mark=at position 0.26 with {\arrow{stealth}},
            mark=at position 0.34 with {\arrow{stealth}},
        }}
    ] table[x=capc,y=pjbittrx_c] {Data/Fig3data1span.txt};
    
    \addplot[c-band,
    mark=*, 
    only marks,
    mark size = 1.5, 
    line width=0.7pt,
    each nth point=4,
    ] table[x=capc,y=pjbittrx_c] {Data/Fig3data1span.txt};

    \addplot[c-band, only marks, 
    mark=square,
    mark size = 1.5, 
    line width=1pt,
    each nth point=4,
    ] table[x=capc,y=pjbitamp_c] {Data/Fig3data1span.txt};

    \node[rotate=-3] at (axis cs: 33.5,35) {\normalsize $E_b \propto 1/T_{\text{tot}} $};

    \draw [-stealth,line width = 1] (axis cs: 39,26)--(axis cs:45,23);
    \draw [-stealth,line width = 1] (axis cs: 39,26)--(axis cs:45,23);
    \node at (axis cs: 42,40) {Increasing $P_i$};

    \node at (axis cs: 38.15, 11) {\footnotesize $-12$};
    \node at (axis cs: 44.55, 10.5) {\footnotesize $-8$};
    \node at (axis cs: 48.75, 10) {\footnotesize $-4$};

    \node at (axis cs: 38.15, 0.035) {\footnotesize $-12$};
    \node at (axis cs: 44.55, 0.03) {\footnotesize $-8$};
    \node at (axis cs: 48.75, 0.0275) {\footnotesize $-4$};
    
    \node at (axis cs: 50.2, 150) {\footnotesize $36$};
    \node at (axis cs: 42, 400) {\footnotesize $40$};

    \draw[black!90,densely dashed] (51,70) rectangle (54,8);





    \end{groupplot}

    \begin{scope}[shift={(0.35,0.7)}]
        \begin{axis}
    [
    width=5.8cm,
    height=3cm,
    grid=major,
    ylabel near ticks,
    xlabel near ticks,
    ylabel shift = -8 pt,
    xlabel shift = -2 pt,
    clip marker paths=true,
    ymode=log,
    xmode=log,
    clip mode=individual,
    label style={font=\footnotesize},
    tick label style={font=\footnotesize},
    legend style={fill opacity=1, draw opacity=1, text opacity=1, at={(0.1,0.5)}, anchor=north west, draw=black, nodes={scale=1, transform shape}, legend columns=1},
    group style={group size=2 by 1,xlabels at=edge bottom,ylabels at=edge left,horizontal sep=1.2cm},
    domain = 30:43,
    xtick = {\empty},
    ytick = {\empty},
    ymin=17,
    ymax=21,
    xmin=34.5,
    xmax=40.5,
    axis background/.style={fill=white}
    ]

    \addplot[black,
    no marks, 
    dashed,
    line width = 0.8pt,
    ]plot {(29*24)/\x};    

    \addplot[l-band,
    no marks,  
    line width=1.2pt,
    postaction={decorate, decoration={
            markings,
            mark=at position 0.15 with {\arrow{stealth}},
        }}
    ] table[x=capc,y=pjbittrx_c] {Data/Fig3dataSMF_3span.txt};
    
    \addplot[l-band,
    mark=*, 
    only marks,
    mark size = 1.5, 
    line width=0.7pt,
    each nth point=4,
    ] table[x=capc,y=pjbittrx_c] {Data/Fig3dataSMF_3span.txt};

    \addplot[black, only marks,mark=square*] coordinates {(37.67,19.02)};
    \node[right] at (axis cs:37.67,19.02) {$T_{\text{tot}}^{\text{max}} \approx E_b^{\text{min}}$};



    \node at (axis cs: 35.7, 19.1) {\footnotesize $9$};
    \node at (axis cs: 37.4, 18.3) {\footnotesize $10$};
    \node at (axis cs: 36.7, 20.4) {\footnotesize $11$};

        \end{axis}
    \end{scope}

    \begin{scope}[shift={(8.5,0.8)}]
        \begin{axis}
    [
    width=4.1cm,
    height=3.7cm,
    grid=major,
    ylabel near ticks,
    xlabel near ticks,
    ylabel shift = -8 pt,
    xlabel shift = -2 pt,
    clip marker paths=true,
    ymode=log,
    xmode=log,
    clip mode=individual,
    label style={font=\footnotesize},
    tick label style={font=\footnotesize},
    legend style={fill opacity=1, draw opacity=1, text opacity=1, at={(0.1,0.5)}, anchor=north west, draw=black, nodes={scale=1, transform shape}, legend columns=1},
    group style={group size=2 by 1,xlabels at=edge bottom,ylabels at=edge left,horizontal sep=1.2cm},
    domain = 50:55,
    xtick = {\empty},
    ytick = {\empty},
    ymin=12.3,
    ymax=17.3,
    xmin=51.3,
    xmax=53.7,
    axis background/.style={fill=white}
    ]
    
    \addplot[black,
    no marks, 
    dashed,
    line width = 0.8pt,
    ]plot {(29*24)/\x};    

    \addplot[c-band,
    no marks,  
    line width=1.2pt,
     postaction={decorate, decoration={
            markings,
            mark=at position 0.03 with {\arrow{stealth}},
        }}
    ] table[x=capc,y=pjbittrx_c] {Data/Fig3data1span.txt};
    
    \addplot[c-band,
    mark=*, 
    only marks,
    mark size = 1.5, 
    line width=0.7pt,
    each nth point=4,
    ] table[x=capc,y=pjbittrx_c] {Data/Fig3data1span.txt};

    \addplot[black, only marks,mark=diamond*,mark size = 3.5pt] coordinates {(52.319,13.39)};
    \node[below] at (axis cs:52.319,13.39) {$E_b^{\text{min}}$};

    \addplot[black, only marks,mark=square*] coordinates {(52.67,16.226)};
    \node[right] at (axis cs:52.67,16.226) {$T_{\text{tot}}^{\text{max}}$};

    \draw [stealth-,line width = 0.7] (rel axis cs: 0.43,0.32)--(rel axis cs:0.43,0.8) node[midway,fill=white,inner sep=1pt,font=\scriptsize,anchor=east] {40.2\%};
    
    \draw [stealth-,line width = 0.7] (rel axis cs: 0.43,0.82)--(rel axis cs:0.54,0.82);

    \node[font=\scriptsize] at (rel axis cs: 0.4,0.91) {-2.2\%};

        \end{axis}
    \end{scope}

\end{tikzpicture}

\caption{Energy-per-bit versus throughput for 200~km C-band transmission and increasing per-channel launch power (solid lines) for (a) SMF and (b) HCF. Insets indicate the minimum energy-per-bit (diamond) and maximum throughput (square).}
\label{fig:3}
\end{figure*}
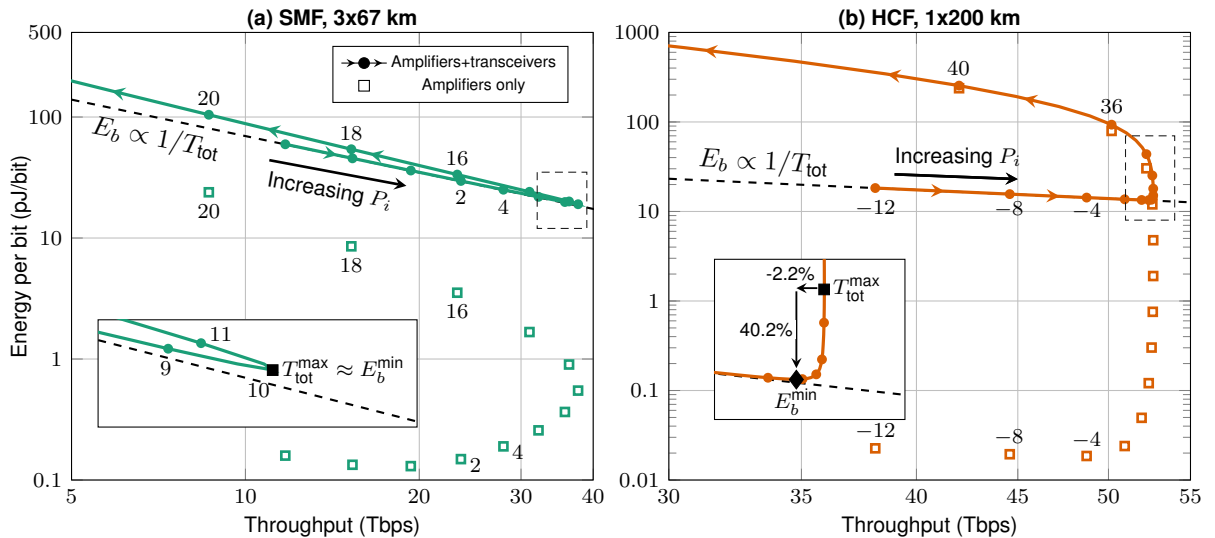

We investigated UWB transmission across the O-band (1265~nm-1355~nm), E-band (1400~nm-1460~nm), S-band (1470~nm-1520~nm), C-band (1530~nm-1565~nm) and L-band (1570~nm-1620~nm) for 105, 59, 63, 29 and 32 channels respectively, shown in Fig.~\ref{fig:1}. We modelled fully-loaded bands with 140~GBd dual-polarisation channels with a 150~GHz spacing and transceiver SNR of 20~dB. The HCF span length was 200~km with an IMI coefficient of -52~dB/km and a nonlinearity coefficient $\gamma$ of $5{\times}10^{-4}$ W$^{-1}$km$^{-1}$ \cite{ospinaLeveragingDigitalSubcarrier2026}. The IMI was assumed to be wavelength-independent, as no measurements are available beyond the C-band. The noise figures of the O-, E-, S-, C- and L-band amplifiers were 5~dB, 6.5~dB, 7~dB, 5~dB and 6~dB, respectively. We used the same constant launch power for all bands and channels.

The total electrical power consumption of a link can be calculated as the sum of the total amplifier power $P^{\text{amp}}_e$ and the total transceiver power $P^{\text{trx}}_e$. The total amplifier power consumption can be calculated as $P^{\text{amp}}_e = (N_{\text{span}}+1)(P_{\text{out}} - P_{\text{in}})/\eta_{\text{PCE}}$ where $P_{\text{in}}$ and $P_{\text{out}}$ are the amplifier input and output powers respectively, $N_{\text{span}}$ is the span number and $\eta_{\text{PCE}}$ is the wallplug PCE. An additional booster amplifier is assumed to achieve total launch powers above 20~dBm. The PCEs for state-of-the-art O-, E-, S-, C- and L-band amplifiers was previously measured in \cite{sohanpalUltraWidebandTransmissionSystems2026}. To determine the PCE for low amplifier output powers, a fitting function was applied to the measured data. 

For the purposes of this work, we assumed the PCE becomes constant at high output powers and that the high-power PCE is the same as that of current amplifiers, for which we use measured values of 0.7\%, 1.3\%, 1.2\%, 6.0\% and 4.4\% for the O-, E-, S-, C- and L-bands. This is an optimistic assumption for high power amplifier PCEs as typically the driver efficiency drops when supplying large currents due to excess thermal dissipation, hence the amplifier power consumption here is underestimated \cite{liangRepeaterPowerConversion2021}. For $N$ channels, the total transceiver power is $N$ times that of a single transceiver. We use a constant 24~W per transceiver for all wavelengths, corresponding to the approximate power consumption of pluggable transceiver modules. The total energy per bit can therefore be calculated as $E_b = (P^{\text{amp}}_e+P^{\text{trx}}_e)/T_{\text{tot}}$.

\section{Energy per bit results}

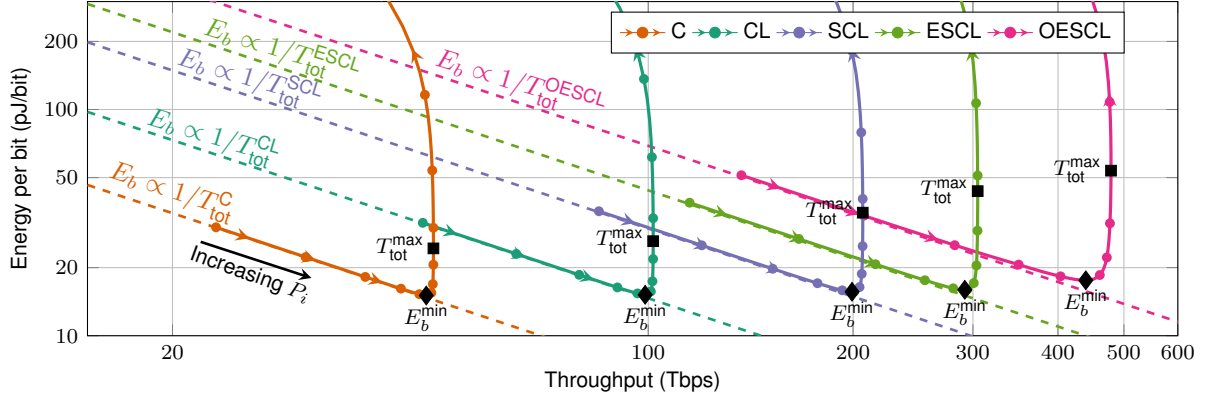
\begin{figure*}[t]
\begin{tikzpicture}[font=\footnotesize]

    \begin{axis}
    [
    width=\linewidth,
    height=6cm,
    ylabel near ticks,
    xlabel near ticks,
    ylabel shift = -2 pt,
    xlabel shift = -2 pt,
    clip marker paths=true,
    ymin=10,
    ymax=300,
    xmin=15,
    xmax=600,
    domain = 15:1000,
    ymode=log,
    xmode=log,
    xlabel=Throughput (Tbps),
    ylabel=Energy per bit (pJ/bit),
    label style={font=\footnotesize},
    tick label style={font=\footnotesize},
    grid=both,
    xtick={20,100, 200, 300, 400, 500, 600},
    xticklabels={$20$,$100$,$200$,$300$,$400$,$500$,$600$},
    ytick = {10,20,50,100,200},
    yticklabels={$10$,$20$,$50$,$100$,$200$},
    legend style={fill opacity=1, draw opacity=1, text opacity=1, at={(0.48,0.97)}, anchor=north west, draw=black, nodes={scale=1, transform shape}, legend columns=5},
    ]



      \addplot[o-band,
    no marks, 
    dashed,
    line width = 1pt,
    ]plot {(288*24)/\x};    

    \addplot[o-band,
    no marks,  
    line width=1.2pt,
    postaction={decorate, decoration={
            markings,
            mark=at position 0.02 with {\arrow{stealth}},
            mark=at position 0.06 with {\arrow{stealth}},
            mark=at position 0.1 with {\arrow{stealth}},
            mark=at position 0.27 with {\arrow{stealth}},
            mark=at position 0.31 with {\arrow{stealth}},
        }}
    ] table[x=capoescl,y=pjbittrx_oescl] {Data/Fig3data5span.txt};
    
    \addplot[o-band,
    mark=*, 
    only marks,
    mark size = 1.5, 
    line width=0.7pt,
    each nth point=4,
    ] table[x=capoescl,y=pjbittrx_oescl] {Data/Fig3data5span.txt};


       \addplot[e-band,
    no marks, 
    dashed,
    line width = 1pt,
    ]plot {(183*24)/\x};    

    \addplot[e-band,
    no marks,  
    line width=1.2pt,
    postaction={decorate, decoration={
            markings,
            mark=at position 0.02 with {\arrow{stealth}},
            mark=at position 0.06 with {\arrow{stealth}},
            mark=at position 0.1 with {\arrow{stealth}},
            mark=at position 0.27 with {\arrow{stealth}},
            mark=at position 0.31 with {\arrow{stealth}},
        }}
    ] table[x=capescl,y=pjbittrx_escl] {Data/Fig3data5span.txt};

    \addplot[e-band,
    mark=*, 
    only marks,
    mark size = 1.5, 
    line width=0.7pt,
    each nth point=4,
    ] table[x=capescl,y=pjbittrx_escl] {Data/Fig3data5span.txt};


    \addplot[s-band,
    no marks, 
    dashed,
    line width = 1pt,
    ]plot {(124*24)/\x};    

    \addplot[s-band,
    no marks,  
    line width=1.2pt,
    postaction={decorate, decoration={
            markings,
            mark=at position 0.02 with {\arrow{stealth}},
            mark=at position 0.06 with {\arrow{stealth}},
            mark=at position 0.1 with {\arrow{stealth}},
            mark=at position 0.27 with {\arrow{stealth}},
            mark=at position 0.31 with {\arrow{stealth}},
        }}
    ] table[x=capscl,y=pjbittrx_scl] {Data/Fig3data5span.txt};
    
    \addplot[s-band,
    mark=*, 
    only marks,
    mark size = 1.5, 
    line width=0.7pt,
    each nth point=4,
    ] table[x=capscl,y=pjbittrx_scl] {Data/Fig3data5span.txt};


    \addplot[l-band,
    no marks, 
    dashed,
    line width = 1pt,
    ]plot {(61*24)/\x};    

    \addplot[l-band,
    no marks,  
    line width=1.2pt,
    postaction={decorate, decoration={
            markings,
            mark=at position 0.02 with {\arrow{stealth}},
            mark=at position 0.06 with {\arrow{stealth}},
            mark=at position 0.1 with {\arrow{stealth}},
            mark=at position 0.27 with {\arrow{stealth}},
            mark=at position 0.31 with {\arrow{stealth}},
        }}
    ] table[x=capcl,y=pjbittrx_cl] {Data/Fig3data5span.txt};
    
    \addplot[l-band,
    mark=*, 
    only marks,
    mark size = 1.5, 
    line width=0.7pt,
    each nth point=4,
    ] table[x=capcl,y=pjbittrx_cl] {Data/Fig3data5span.txt};


    \addplot[c-band,
    no marks, 
    dashed,
    line width = 1pt,
    ]plot {(29*24)/\x};    

    \addplot[c-band,
    no marks,  
    line width=1.2pt,
    postaction={decorate, decoration={
            markings,
            mark=at position 0.02 with {\arrow{stealth}},
            mark=at position 0.06 with {\arrow{stealth}},
            mark=at position 0.1 with {\arrow{stealth}},
            mark=at position 0.27 with {\arrow{stealth}},
            mark=at position 0.31 with {\arrow{stealth}},
        }}
    ] table[x=capc,y=pjbittrx_c] {Data/Fig3data5span.txt};
    
    \addplot[c-band,
    mark=*, 
    only marks,
    mark size = 1.5, 
    line width=0.7pt,
    each nth point=4,
    ] table[x=capc,y=pjbittrx_c] {Data/Fig3data5span.txt};

    \node[rotate=-18,c-band] at (axis cs: 20,44) {\normalsize $E_b \propto 1/T_{\text{tot}}^{\text{C}} $};
    \node[rotate=-17,l-band] at (axis cs: 23,76) {\normalsize $E_b \propto 1/T_{\text{tot}}^{\text{CL}} $};
    \node[rotate=-17,s-band] at (axis cs: 26,138) {\normalsize $E_b \propto 1/T_{\text{tot}}^{\text{SCL}} $};
    \node[rotate=-17,e-band] at (axis cs: 29,183) {\normalsize $E_b \propto 1/T_{\text{tot}}^{\text{ESCL}} $};
    \node[rotate=-17,o-band] at (axis cs: 65,130) {\normalsize $E_b \propto 1/T_{\text{tot}}^{\text{OESCL}} $};

    \addplot[black, only marks,mark=square*] coordinates {(48.35,24.327)};
    \addplot[black, only marks,mark=square*] coordinates {(101.69,26.18)};
    \addplot[black, only marks,mark=square*] coordinates {(206.64,35)};
    \addplot[black, only marks,mark=square*] coordinates {(304.87,43.56)};
    \addplot[black, only marks,mark=square*] coordinates {(478.62,53.66)};

    \addplot[black, only marks,mark=diamond*,mark size = 3.5pt] coordinates {(47.19,15.06)};
    \addplot[black, only marks,mark=diamond*,mark size = 3.5pt] coordinates {(98.96,15.17)};
    \addplot[black, only marks,mark=diamond*,mark size = 3.5pt] coordinates {(199.32,15.64)};
    \addplot[black, only marks,mark=diamond*,mark size = 3.5pt] coordinates {(291.84,15.962)};
    \addplot[black, only marks,mark=diamond*,mark size = 3.5pt] coordinates {(439.49,17.5808)};

    \node[below] at (axis cs:47.19,15.06) {$E_b^{\text{min}}$};
    \node[below] at (axis cs:98.96,15.17) {$E_b^{\text{min}}$};
    \node[below] at (axis cs:199.32,15.64) {$E_b^{\text{min}}$};
    \node[below] at (axis cs:291.84,15.962) {$E_b^{\text{min}}$};
    \node[below] at (axis cs:439.49,17.5808) {$E_b^{\text{min}}$};

    \node[left] at (axis cs:48.35,24.327) {$T_{\text{tot}}^{\text{max}}$};
    \node[left] at (axis cs:101.69,26.18) {$T_{\text{tot}}^{\text{max}}$};
    \node[left] at (axis cs:206.64,35) {$T_{\text{tot}}^{\text{max}}$};
    \node[left] at (axis cs:304.87,43.56) {$T_{\text{tot}}^{\text{max}}$};
    \node[left] at (axis cs:478.62,53.66) {$T_{\text{tot}}^{\text{max}}$};

    \draw [-stealth,line width = 1] (axis cs: 22,26)--(axis cs:32,18);
    \node[rotate=-20] at (axis cs: 26,18.5) {Increasing $P_i$};

    \addlegendimage{c-band,mark=*,
    postaction={decorate, decoration={
            markings,
            mark=at position 0.3 with {\arrow{stealth}},
            mark=at position 0.9 with {\arrow{stealth}}
    }}}
    \addlegendimage{l-band,mark=*,
    postaction={decorate, decoration={
            markings,
            mark=at position 0.3 with {\arrow{stealth}},
            mark=at position 0.9 with {\arrow{stealth}}
    }}}
    \addlegendimage{s-band,mark=*,
    postaction={decorate, decoration={
            markings,
            mark=at position 0.3 with {\arrow{stealth}},
            mark=at position 0.9 with {\arrow{stealth}}
    }}}
    \addlegendimage{e-band,mark=*,
    postaction={decorate, decoration={
            markings,
            mark=at position 0.3 with {\arrow{stealth}},
            mark=at position 0.9 with {\arrow{stealth}}
    }}}
    \addlegendimage{o-band,mark=*,
    postaction={decorate, decoration={
            markings,
            mark=at position 0.3 with {\arrow{stealth}},
            mark=at position 0.9 with {\arrow{stealth}}
    }}}

    \legend{,,,,,,,,,,,,,,,,,,,,,,,,,C,CL,SCL,ESCL,OESCL}


    \end{axis}

\end{tikzpicture}

\caption{Total energy per bit versus throughput for the C-, CL-, SCL-, ESCL- and OESCL-bands for increasing launch power for 5x200~km HCF (1000~km total). Markers indicate minimum energy per bit $E_b^{\text{min}}$ and maximum throughput $T_{\text{tot}}^{\text{max}}$.}
\label{fig:4}
\end{figure*}
\vspace{-0.3em}
Using the aforementioned model in the C-band, the total throughput versus power per channel is shown in Fig.~\ref{fig:2} for 3x67~km low-loss SMF (neglecting ISRS) and 1x200~km HCF. The maximum throughput $T_{\text{tot}}^{\text{max}}$ of the SMF system is strongly limited by Kerr nonlinearity, leading to a drop in throughput once the launch power exceeds 10~dBm per channel (24.5~dBm total).

Fig.~\ref{fig:3}a shows the calculated energy per bit $E_b$ versus throughput for C-band transmission in a 3x67~km SMF link for varying channel launch powers. At low launch powers (e.g. \textless0~dBm per channel), the total amplifier electrical power consumption is much smaller than that of the transceivers, $P^{\text{amp}}_e \ll P^{\text{trx}}_e$, thus the energy per bit is determined by $E_b = P^{\text{trx}}_e/T_{\text{tot}}$. As $P^{\text{trx}}_e$ is independent of launch power, $E_b \propto 1/T_{\text{tot}}$. As the launch power increases, $E_b$ decreases proportionally to the increase in throughput. Beyond the optimum launch power ($T_{\text{tot}}^{\text{max}}$) the throughput decreases rapidly and $E_b$ increases (but still nearly proportionally to $1/T_{\text{tot}}$). Thus, the SMF link power consumption is dominated by the transceivers and the amplifier power consumption is negligible for few-span SMF transmission systems.

For a HCF link, the throughput in Fig.~\ref{fig:2} is limited by transceiver SNR and exhibits a throughput plateau that extends by more than 30~dB in launch power. The HCF energy per bit is shown in Fig.~\ref{fig:3}b. At low launch powers, the HCF link $E_b$ is also $P^{\text{trx}}_e$-dominated and follows the $1/T_{\text{tot}}$ scaling. However, as the launch power approaches optimum, the amplifier power consumption becomes comparable to (and eventually exceeds) that of the transceiver, $P^{\text{amp}}_e \sim P^{\text{trx}}_e$, leading to a sharp increase in $E_b$. This occurs due to two factors. Firstly, the launch power near the maximum throughput grows to very large values, exceeding 20~dBm per channel (35~dBm total) due to the low Kerr nonlinearity. Secondly, the throughput of HCF links is strongly curtailed by the transceiver SNR (and to a lesser extent IMI), meaning large increases in launch power lead to only negligible throughput gain (hence $P^{\text{amp}}_e$ grows faster than $T_{\text{tot}}$). 

This leads to an interesting conclusion - unlike SMF links, in HCF the launch power for maximum throughput $T_{\text{tot}}^{\text{max}}$ is no longer similar to the launch power for minimum energy per bit $E_b^{\text{min}}$ (Fig.~\ref{fig:3} insets) even for a single 200~km span. For a 1000~km C-band link, operating at $E_b^{\text{min}}$ leads to a 40.2\% improvement in overall energy per bit (41.5\% in total power) for a 2.2\% penalty in throughput compared to at $T_{\text{tot}}^{\text{max}}$. We extended this analysis to multi-band systems, shown in Fig.~\ref{fig:4}. Table~\ref{fig:tab1} shows the throughput reduction and the $E_b$ improvement for each scenario when operating at $T_{\text{tot}}^{\text{max}}$ and $E_b^{\text{min}}$ total launch powers, showing more than 71.9\% reduction in total power consumption for an 8.5\% throughput penalty in OESCL-band systems. In all UWB scenarios, $E_b^{\text{min}}$ occurs around 7~dBm per channel (22~dBm total for C-band) and $T_{\text{tot}}^{\text{max}}$ occurs around 22~dBm per channel (37~dBm total for C-band), thus significant reduction in power consumption and amplifier requirements are achievable by appropriate launch power selection, alleviating the need for high-power amplifiers in transceiver SNR-limited scenarios.

\begin{table}[]
\resizebox{\columnwidth}{!}{%
\begin{tabular}{@{}cccccc@{}}
\toprule
 & \textbf{C} & \textbf{CL} & \textbf{SCL} & \textbf{ESCL} & \textbf{OESCL} \\ \midrule
Throughput change (\%) & -2.2 & -2.6 & -3.5 & -4.6 & -8.5 \\
$E_b$ reduction (\%) & 40.2 & 45.1 & 52.9 & 62.2 & 69.3 \\ 
Total power savings (\%) & 41.5 & 46.5 & 54.6 & 64.0 & 71.9 \\ \bottomrule
\end{tabular}
}

\caption{Throughput change and $E_b$ reduction when operating at $E_b^{\text{min}}$ instead of $T_{\text{tot}}^{\text{max}}$ per-channel launch power at 1000~km.}
\label{fig:tab1}
\end{table}

\vspace{-0.3em}

\section{Conclusion}

We investigated the optimal launch powers for energy-efficient HCF transmission systems. We observe a significant difference in energy per bit when operating at launch powers achieving either maximum-throughput or minimum-energy-per-bit. For 1000~km C-band transmission, minimum-energy-per-bit operation leads to a 41.5~\% reduction in total power consumption for a 2\% throughput penalty compared to maximum throughput, enabling energy-efficient HCF operation.

\clearpage
\section{Acknowledgements}
This work was supported by EPSRC grant EP/R035342/1 TRANSNET. Eric Sillekens was supported by the Department for Science, 
Innovation and Technology and the Royal Academy of Engineering under the Research Fellowship scheme. Polina Bayvel is supported under a Royal Society Research Professorship.

\printbibliography

\vspace{-4mm}

\end{document}